\documentclass{article}
\usepackage{amsmath}
\usepackage{amsthm}
\usepackage{amsfonts}
\usepackage{amssymb}
\usepackage{amscd}
\usepackage[mathscr]{eucal}
\usepackage{mathtools}
\usepackage{fullpage}
\usepackage{color}
\usepackage{float}
\usepackage{graphicx}
\usepackage{natbib}
\usepackage[hidelinks]{hyperref}

\newcommand*{\myabstract}[1]{ \begin{abstract} #1 \end{abstract} }
\begin{document}
\title{\bf Guided waves as superposition of body waves}
\author{
David R. Dalton%
\footnote{Department of Earth Sciences, Memorial University of Newfoundland,  Canada; 
{\tt dalton.nfld@gmail.com}}\,,
Michael A. Slawinski%
\footnote{Department of Earth Sciences, 
Memorial University of Newfoundland, Canada; 
{\tt mslawins@mac.com}}\,,
Theodore Stanoev%
\footnote{Department of Earth Sciences, 
Memorial University of Newfoundland, Canada; 
{\tt theodore.stanoev@gmail.com}}
\date{}
}
\maketitle
\myabstract{%
We illustrate properties of guided waves in terms of a superposition of body waves.
In particular, we consider the Love and $SH$ waves.
Body-wave propagation at postcritical angles---required for a total reflection---results in the speed of the Love wave being between the speeds of the $SH$ waves in the layer and in the halfspace.
A finite wavelength of the $SH$ waves---required for constructive interference---results in a limited number of modes of the Love wave.
Each mode exhibits a discrete frequency and propagation speed; the fundamental mode has the lowest frequency and the highest speed.}
\section{Introduction}
Let us consider Love wave and the $SH$ waves to examine the concept of a guided wave within a layer as an interference of body waves therein.
In the $x_1x_3$-plane, the nonzero component of the displacement vector of the Love wave is \citep[e.g.,][Section~6.3]{Slawinski}
\begin{align*}
u_2^\ell(x_1,x_3,t) 
=&\,\,
C_1\exp{\left(-\iota\,\kappa\,s_\ell\,x_3\right)}\exp\left[\iota\,(\kappa\,x_1-\omega\,t)\right]\\
&+C_2\exp{\left(\iota\,\kappa\,s_\ell\,x_3\right)}\exp\left[\iota\,(\kappa\,x_1-\omega\,t)\right]\,,
\end{align*}
where $s_\ell:=\sqrt{(v/\beta_\ell)^2-1}$\,, with $v$ being the speed of the Love wave and $\beta_\ell$ the speed of the $SH$ wave; $\omega$ and $\kappa$ are the temporal and spatial frequencies, related by $\kappa=\omega/v$\,.
The~$SH$\index{SH@$SH$ wave}\index{wave!SH@$SH$} waves travel obliquely in the $x_1x_3$-plane; different signs in front of $x_3$ mean that one wave travels upwards and the other downwards.
Their wave vectors\index{wave!vector} are~${\bf k}_{\pm}:=(\kappa, 0, \pm\,\kappa\,s_\ell)$\,.
Considering their magnitudes,
\begin{equation*}
\|{\bf k}_\pm\|=\sqrt{\kappa^2+(\kappa\,s_\ell)^2}\,,
\end{equation*}
we have
\begin{equation*}
\|{\bf k}_\pm\|=\kappa\sqrt{1+(s_\ell)^2}=\kappa\,\dfrac{v}{\beta_\ell}\,;
\end{equation*}
from which it follows that  
\begin{equation}
\label{eq:LoveRestrict}
\dfrac{\beta_\ell}{v}=\dfrac{\kappa}{\|\bf k_\pm\|}=\sin \theta\,,
\end{equation}
where~$\theta$ is the angle between~${\bf k}_{\pm}$ and the~$x_3$-axis.
Thus,~$\theta$ is the angle between the~$x_3$-axis and a wavefront normal, which means that---exhibiting opposite signs---it is the propagation direction of upward and downward wavefronts.
\section{Total reflection}
A necessary condition for the existence of a guided wave is a {\sl total reflection} on either side of the layer; the energy must remain within a layer.
For the Love\index{Love wave}\index{wave!Love} waves, this is tantamount to no transmission of the $SH$\index{SH@$SH$ wave}\index{wave!SH@$SH$} waves through the surface or the interface.
The former is ensured by the assumption of vacuum above the surface; hence, total reflection\index{total reflection} occurs for all propagation angles,~$\theta$\,.
The latter requires $\beta_\ell<\beta_h$\,, where $\beta_h$ is the speed of the $SH$\index{SH@$SH$ wave}\index{wave!SH@$SH$} wave within the halfspace.
This inequality results in the existence of a critical angle\index{critical angle!guided wave},~$\theta_c=\arcsin(\beta_\ell/\beta_h)$\,, which is required for a propagation at postcritical angles, $\theta>\theta_c$\,.

In view of expression~(\ref{eq:LoveRestrict}), the lower limit of $v$ is $\beta_\ell$\,, for which $\sin\theta=\beta_\ell/\beta_\ell=1$\,; hence, $\theta=\pi/2$\,.
It corresponds to the $SH$\index{SH@$SH$ wave}\index{wave!SH@$SH$} waves that propagate parallel to the~$x_1$-axis, and can be viewed as the Love\index{Love wave}\index{wave!Love} wave.

The upper limit, $v=\beta_h$\,, is a consequence of the critical angle, for which $\sin\theta_c=\beta_\ell/\beta_h$\,.
If $\beta_h\to\infty$\,---which corresponds to a rigid halfspace---\,$\theta_c\to 0$\,; hence, the $SH$~waves within the layer can propagate nearly perpendicularly to the interface and still exhibit a total reflection.
This means that $v\to\infty$\,, as can be also inferred from Figure~\ref{fig:GuideTwo}.

These limits, $\beta_\ell<v<\beta_h$\,, are a consequence of total reflection.
Also, the upper limit needs to be introduced to ensure an exponential amplitude decay in the halfspace \citep[e.g.,][Section~6.3.2]{Slawinski}.
\section{Constructive interference}
\begin{figure}
\begin{center}
\includegraphics[width=11cm]{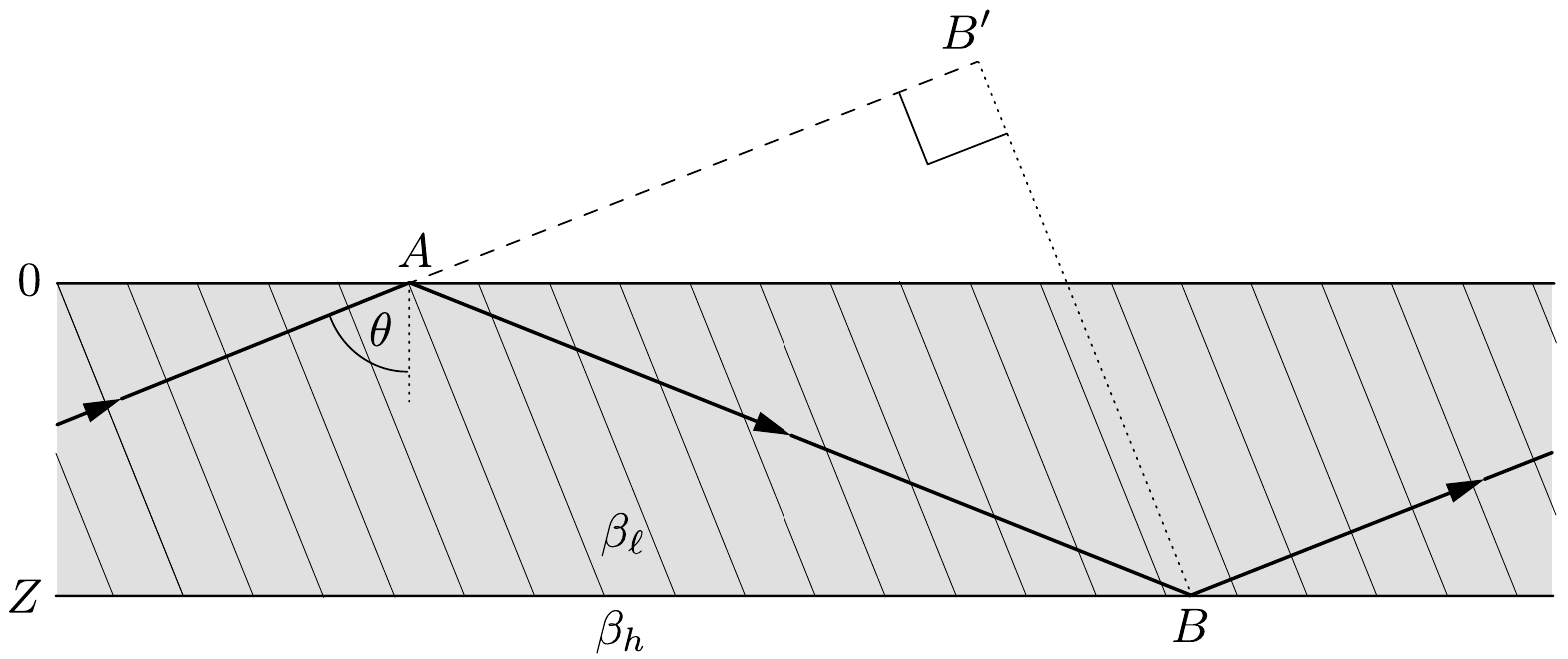}
\end{center}
\caption{\small{Constructive interference for Love\index{Love wave}\index{wave!Love} wave\,: The $SH$\index{SH@$SH$ wave}\index{wave!SH@$SH$} wave reflected twice reproduces itself, and, hence, the nonreflected and twice-reflected $SH$ wavefronts coincide.
Herein, $\|AB\|=a\,\lambda$\,, and $\|AB'\|=b\,\lambda$, where $(a-b)\in{\mathbb N}$\,.}}
\label{fig:GuideOne}
\end{figure}

Guided waves---as superpositions of body waves---require a {\sl constructive interference}\index{constructive interference!guided wave} of body waves.
A necessary condition of such an interference\index{constructive interference!guided wave} is the same phase among the wavefronts of parallel rays.
In Figure~\ref{fig:GuideOne}, this condition means that the difference between $\|AB\|$ and $\|AB'\|$ must be equal to a positive-integer multiple of the wavelength,~$\lambda$\,, taking into account the phase shift due to reflection.
A reflection at the surface results in no phase shift \citep[Sections~5.4 and 10.3.1]{Udias}, and the $SH$-wave postcritical phase shift at the elastic halfspace is presented by \citet[equation~(5.74)]{Udias}.

To illustrate the constructive interference---without discussing the phase shift as a function of the incidence angle---let us consider an elastic layer above a rigid halfspace, on which a transverse wave undergoes a phase change of $\pi$~radians for any angle.
In such a case, the propagation angle is \citep[e.g.,][Section~7.1]{SalehTeich} 
\begin{equation}
\label{eq:ConstInterQ}
\theta_n=\arcsin\left(n\frac{\lambda}{Z}\right)\,,\qquad n=1,2,\ldots\,,
\end{equation}
where $\lambda$ is the wavelength of the $SH$ wave, $Z$ is the layer thickness and $n$ is a mode of the guided wave; $n=1$ is the fundamental mode\index{fundamental mode}.

\begin{figure}
\begin{center}
\includegraphics[width=11cm]{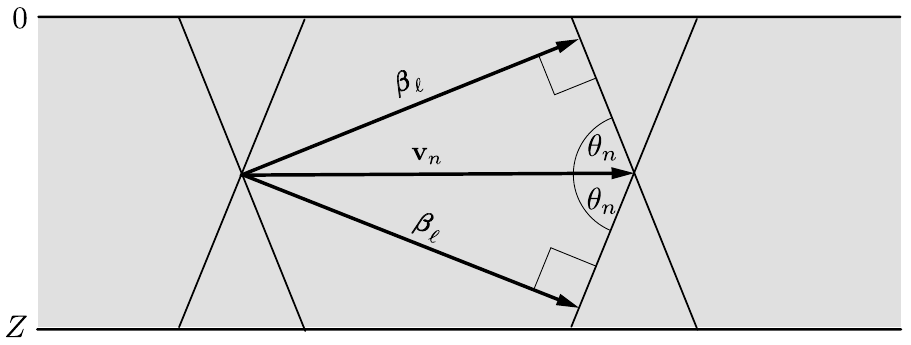}
\end{center}
\caption{\small{Constructive interference for Love\index{Love wave}\index{wave!Love} wave\,: The upgoing and downgoing $SH$\index{SH@$SH$ wave}\index{wave!SH@$SH$} wavefronts at two instants; their speed,~$\|\pmb{\beta}_\ell\|$\,, remains constant---regardless of the wavefront orientation---but the Love-wave speed,~$\|{\bf v}_n\|$\,, whose direction, ${\bf v}_n$\,, remains constant, increases as $\theta_n$ decreases.}}
\label{fig:GuideTwo}
\end{figure}

Thus---as a consequence of constructive interference\index{constructive interference!guided wave}---for a given $SH$\index{SH@$SH$ wave}\index{wave!SH@$SH$} wavelength and layer thickness, the propagation angles, $\theta_n$\,, form a set of discrete values; each $n$ corresponds to a mode of the guided wave.
As illustrated in Figure~\ref{fig:GuideTwo}, each mode has its propagation speed, which---in accordance with expression~(\ref{eq:LoveRestrict})---is
\begin{equation}
\label{eq:vnspeed}
v_n=\dfrac{\beta_\ell}{\sin\theta_n}\,,
\end{equation}
where, as a consequence of total reflection\index{total reflection}, $\theta_n\in(\theta_1,\pi/2)$\,, where $\theta_1>\theta_c$\,.
The specific value of $\theta_1$ depends on~$Z$ and $\lambda$\,; it corresponds to the first postcritical value for which $\|AB\|-\|AB'\|=2\,\|AB\|\cos^2\theta=2\,Z\,\cos\theta$ is a multiple integer of~$\lambda$\,.

Examining Figure~\ref{fig:GuideTwo}, we distinguish the upgoing and downgoing wavefronts, which compose the guided wave.
Its longest permissible wavelength is twice the layer thickness, $\lambda_1=2Z$\,, which corresponds to the fundamental mode\index{fundamental mode}; $\lambda_2=Z$\,, $\lambda_3=2Z/3$\,, and, in general, $\lambda_n=2Z/n$\,.
\section{Frequencies of body and guided waves}
$\lambda$\,, referred to in the caption of Figure~\ref{fig:GuideOne} and used in expression~(\ref{eq:ConstInterQ}), corresponds to the $SH$\index{SH@$SH$ wave}\index{wave!SH@$SH$} wave; $\lambda_n$\,, where $n=1,2,\ldots$\,, corresponds to the guided wave.
They are related by the propagation angle,~$\theta_n$\,, and by the layer thickness,~$Z$\,.

The radial frequency of a monochromatic $SH$\index{SH@$SH$ wave}\index{wave!SH@$SH$} wave is constant,~$\omega=2\pi\,\beta_\ell/\lambda$\,.
The radial frequencies of the Love\index{Love wave}\index{wave!Love} wave are distinct for distinct modes, $\omega_n=2\pi\,v_n/\lambda_n$\,.
For a given model, $\beta_\ell$\,, $\beta_h$ and $Z$\,, the relations between $\omega$ and $\omega_n$\,, as well as among $\omega_n$\,, where $n=1,2,\ldots$\,, are functions of $n$ and $\theta_n$\,; explicitly, $\omega_n=n\,\pi\,\beta_\ell/(Z\,\sin\theta_n)$\,, and, in general, its behaviour as a function of~$n$ cannot be examined analytically.
However, in an elastic layer above a rigid halfspace, in accordance with expression~(\ref{eq:ConstInterQ}),
\begin{equation}
\label{eq:omegan}
\omega_n=n\,\pi\,\dfrac{\beta_\ell}{Z\,\sin\theta_n}=\pi\,\dfrac{\beta_\ell}{\lambda}=\dfrac{\omega}{2}\,,
\end{equation}
which is constant for all modes, and depends only on the radial frequency of the $SH$\index{SH@$SH$ wave}\index{wave!SH@$SH$} wave.

The constructive interference\index{constructive interference!guided wave}, illustrated in Figure~\ref{fig:GuideOne}, requires that
\begin{equation*}
\|AB\|-\|AB'\|=a\,\lambda-b\,\lambda=(a-b)\,\lambda\,,
\end{equation*}
where---in contrast to $a=\|AB\|/\lambda$ and $b=\|AB'\|/\lambda$---\,$a-b$ is a positive integer; $\lambda$ is the $SH$ wavelength.
Following trigonometric relations, we write
\begin{align*}
\|AB\|-\|AB'\|=&~\|AB\|-\|AB\|\cos(\pi-2\theta)=\|AB\|\left(1-\cos(\pi-2\theta)\right)\\=&~2\,\|AB\|\cos^2\theta\,.	
\end{align*}
Since $\|AB\|=Z/\cos\theta$\,, where $\theta$ is the $SH$\index{SH@$SH$ wave}\index{wave!SH@$SH$}-wave propagation angle, it follows that $2\|AB\|\cos^2\theta=2Z\cos\theta$\,, and the constructive interference\index{constructive interference!guided wave} requires that $2\,Z\cos\theta=(a-b)\,\lambda$\,, where ${(a-b)\in{\mathbb N}}$\,; in other words,
\begin{equation*}
\cos\theta=\dfrac{a-b}{2\,Z}\,\lambda\,,
\end{equation*}
where\index{critical angle!guided wave} $\theta\geqslant\theta_c$\,, to ensure the total reflection\index{total reflection}, and $(a-b)\,\lambda\leqslant2Z$\,, for $\theta\in\mathbb R$\,.

Using this result and the inverse trigonometric function, we write the first equality of expression~(\ref{eq:omegan}) as
\begin{equation}
\label{eq:omega2}
\omega_n=n\,\pi\,\dfrac{\beta_\ell}{Z\,\sqrt{1-\left(\dfrac{a_n-b_n}{2\,Z}\right)^2\,\lambda^2}}\,,
\end{equation}
which corresponds only to a given value of $n$ and, hence, of $\theta_n$\,, since $a_n-b_n$ changes with the propagation angle, and needs to be restricted to integer values for each~$n$\,.

Following expression~(\ref{eq:omega2}), we obtain
\begin{equation}
\label{eq:omega/omega}
\dfrac{\omega_n}{\omega_{n+1}}=
\dfrac{n}{n+1}
\dfrac{\sqrt{1-\left(\dfrac{a_{n+1}-b_{n+1}}{2\,Z}\right)^2\,\lambda^2}}{\sqrt{1-\left(\dfrac{a_n-b_n}{2\,Z}\right)^2\,\lambda^2}}
\,.
\end{equation}
Since $\theta_{n+1}>\theta_n$\,, examining Figure~\ref{fig:GuideOne} and considering given values of $\lambda$ and $Z$\,, we see that---as $\theta$ increases---$\|AB\|-\|AB'\|$ decreases.
Hence, $(a_n-b_n)>(a_{n+1}-b_{n+1})$\,, and the root in the numerator is greater than in the denominator.
Consequently, the ratio of roots is greater than unity.
However, $n/(n+1)<1$\,.

We cannot, in general, determine analytically if the radial frequency of the $n$th mode is higher or lower than the frequency of the $n+1$ mode.
To determine it, we need not only to specify $Z$ and the model parameters, which result in~$\theta_c$\,, but also $\lambda$ and $n$\,, to obtain $\theta_n$ and $\theta_{n+1}$\,, with integer values of $\|AB\|-\|AB'\|$\,.
\section{Numerical example}
To obtain specific values, we let $Z=1000$\,, $\beta_\ell=2000$\,, $\beta_h=3000$ and $\lambda=50$\index{SH@$SH$ wave}\index{wave!SH@$SH$}\,, which means that $\theta_c\approx 0.73$\,, in radians, and ${\omega=2\pi\,\beta_\ell/\lambda\approx 251}$\,.
For the guided wave\index{guided wave}\index{wave!guided}, in accordance with Figure~\ref{fig:GuideOne}, we obtain---numerically---\,${\theta_1\approx 0.76}$\,, which corresponds to $(a_1-b_1)=29$\,.
To include higher modes, using expression~(\ref{eq:omega/omega}), we obtain $\omega_1/\omega_2\approx 0.52$\,, $\omega_2/\omega_3\approx 0.69$ and $\omega_3/\omega_4\approx 0.77$\,, which corresponds to, respectively, $(a_{n+1}-b_{n+1})=29-n=28$\,, $27$ and $26$\,, and to $\omega_2=17.60$\,, $\omega_3=25.55$ and $\omega_4=33.07$\,.

\begin{figure}
\begin{center}
\includegraphics[width=8cm]{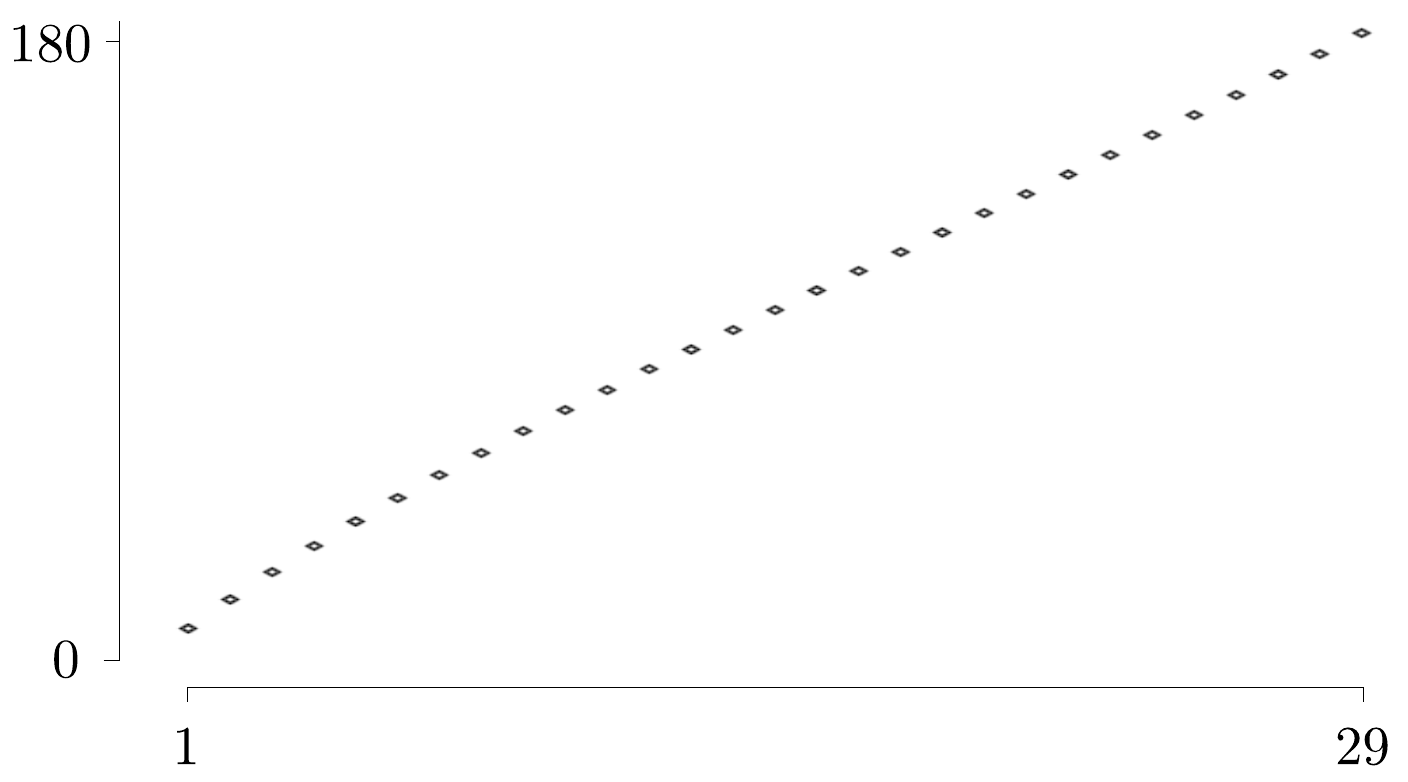}
\quad
\includegraphics[width=8cm]{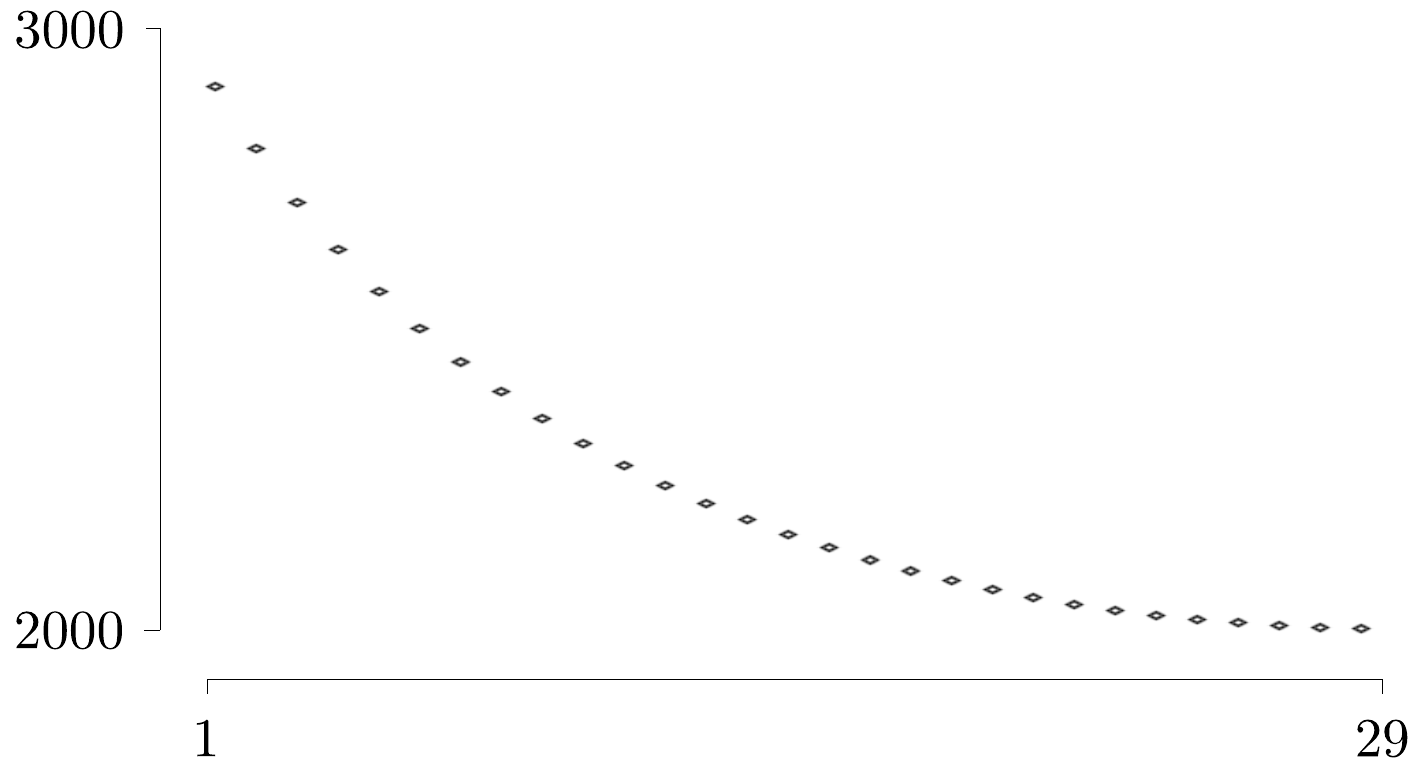}
\end{center}
\caption{\small{Frequencies, $\omega_n$\,, and speeds, $v_n$\,, of Love-wave modes,~$n=1,\,\ldots\,,29$}}
\label{fig:GuideThree}
\end{figure}

We might infer that the \index{fundamental mode} Love-wave\index{Love wave}\index{wave!Love} fundamental mode,~$n=1$\,, exhibits the lowest radial frequency---which, following expression~(\ref{eq:omega2}), is~$9.123$---and that the frequency increases monotonically with $n$\,.
The highest allowable mode corresponds to $n=29$\,, since, for that value, $(a_n-b_n)=1$\,.
For this mode, $\omega_{29}\approx 182.27$\,; also, $\omega_{28}/\omega_{29} \approx 0.966$\,.
Frequencies of distinct modes are shown in the left-hand plot of Figure~\ref{fig:GuideThree}.

Examining expression~(\ref{eq:omega/omega}), in view of these results, we conclude that---as $n$ increases---both $n/(n+1)$ and the ratio of roots tend to unity; the former from below, the latter from above.
The ratio of successive frequencies approaches the ratio of successive overtones for a vibrating string, $\tfrac{1}{2}\,,\tfrac{2}{3}\,,\tfrac{3}{4}\,,\ldots\,, \tfrac{28}{29}$\,.

Furthermore, using expression~(\ref{eq:vnspeed}) and the computed values of~$\theta_n$\,, we can obtain the corresponding propagation speeds of the Love-wave\index{Love wave}\index{wave!Love} modes.
For the fundamental mode,
\begin{equation*}
v_1=\dfrac{\beta_\ell}{\sin\theta_1}=\dfrac{2000}{\sin(0.76)}=2903.09\,,
\end{equation*}
 which is the highest speed of this Love wave\index{Love wave}\index{wave!Love}; it is smaller than $\beta_h=3000$\,, as required.
The lowest speed corresponds to $\theta_{29}=1.55$\,, which is nearly~$\pi/2$\,; hence, the $SH$\index{SH@$SH$ wave}\index{wave!SH@$SH$} waves propagate almost parallel to the layer.
The speed of the resulting Love wave\index{Love wave}\index{wave!Love} is $v_{29}=\beta_\ell/\sin\theta_{29}=2000.63$\,, which is greater than $\beta_\ell=2000$\,, as required.
Speeds of distinct modes are shown in right-hand plot of Figure~\ref{fig:GuideThree}.
\section{Conclusions}
Superposition of body waves allows us to examine several properties of guided waves.
Body-wave propagation at postcritical angles---required for a total reflection---results in the speed of the Love wave being between the speeds of the $SH$ waves in the layer and in the halfspace.
A finite wavelength of the $SH$ waves---required for constructive interference---results in a limited number of modes of the Love wave.
Each mode exhibits a discrete frequency and propagation speed; the first mode has the lowest frequency and the highest speed.
\section*{Acknowledgments}
We wish to acknowledge the graphic support of Elena Patarini.
This research was performed in the context of The Geomechanics Project supported by Husky Energy. 
Also, this research was partially supported by the Natural Sciences and Engineering Research Council of Canada, grant 202259.
\bibliographystyle{apa}
\bibliography{DSSS_arXiv}
\end{document}